\documentclass{kapproc} 

\usepackage{procps} 

\usepackage[dvips]{graphicx}

\upperandlowercase

\kluwerbib 

\newcommand{\HI}{{\protect\sc{HI}}}
\newcommand{\msun}{$M_\odot$}
\newcommand{\etal}{{et~al.}}
\newcommand{\mhi}{$M_{HI}$}
\newcommand{\kms}{km\,s$^{-1}$}

\begin{document}



\articletitle[]{Studying Galaxy Formation in Loose Galaxy Groups}











--------------

%
%

--------------

\author{D.J. Pisano\altaffilmark{1,2}, D.G. Barnes\altaffilmark{3}, 
        B.K. Gibson\altaffilmark{4}, L. Staveley-Smith\altaffilmark{5},
        K.C. Freeman\altaffilmark{6}, V.A. Kilborn\altaffilmark{4}}

\altaffiltext{1}{Naval Research Laboratory, 4555 Overlook Ave. SW,
Washington, DC 20375 USA; pisano@nrl.navy.mil}
\altaffiltext{2}{National Research Council Research Fellow}
\altaffiltext{3}{School of Physics, University of Melbourne, Victoria 
3010, Australia; dbarnes@astro.ph.unimelb.edu.au}
\altaffiltext{4}{Centre for Astrophysics \& Supercomputing, Swinburne 
University, Hawthorn, Victoria 3122, Australia; bgibson@swin.edu.au, 
vkilborn@astro.swin.edu.au}
\altaffiltext{5}{Australia Telescope National Facility, P.O. Box 76, Epping 
NSW 1710, Australia; Lister.Staveley-Smith@atnf.csiro.au}
\altaffiltext{6}{RSAA, Mount Stromlo Observatory, Cotter Road, Weston, ACT 
2611, Australia; kcf@mso.anu.edu.au}

--------------

%



\begin{abstract}
We present the results of our HI survey of six loose groups of galaxies
analogous to the Local Group.  The survey was conducted using the Parkes
telescope and the Australia Telescope Compact Array to produce a census
of all the gas-rich galaxies and analogs to the high-velocity clouds (HVCs)
within these groups down to M$_{HI}$< 10$^7~$M$_\odot$ as a test of models of
galaxy formation.  We present the HI mass function and halo mass function
of the loose groups and show that they are consistent with those of the Local 
Group.  We discuss the possible role of HVCs in solving the ``missing 
satellite'' problem and discuss the implications of our observations for models
of galaxy formation.
\end{abstract}



\section{Introduction}

The majority of ``island universes'', 60\%, reside in galaxy groups
(Tully 1987), including the Milky Way.  Loose groups of galaxies,
like our Local Group,  are collections of a few large, bright galaxies
and tens  of smaller, fainter galaxies.  The large galaxies are
typically separated by a few hundred kiloparsecs from each other and
spread over an extent of approximately  a megaparsec.  They represent
the most diffuse components of large scale structure and a laboratory
for studying galaxy formation.  Loose groups are almost  certainly
still in the process of collapsing (Zabludoff \& Mulchaey 1998)  and
they may also contain the gaseous remnants of galaxy formation in the
form of the high-velocity clouds (HVCs; Blitz \etal\ 1999).   They
also illustrate one of the major challenges to current models of cold
dark matter (CDM) galaxy formation.  CDM models predict that the Local
Group should contain $\sim$300 low mass dark halos, while there are
only $\sim$20 luminous dwarf galaxies known.  While this may imply
that we lack a complete census of the luminous galaxies in the Local
Group, it may also be  uniquely deficient in dwarf galaxies.  Or,
perhaps, the HVCs may  populate these dark matter halos and solve the
``missing satellite''  problem.  In order to address this question and
to better understand  the properties of loose groups, we have
conducted an \HI\  survey of six groups analogous to the Local Group.

\section{Observations}
We selected our groups from the optical  catalog of Garcia (1993).
The resulting groups are  between 10.6 and 13.4 Mpc distant, contain
between 3-9 bright  galaxies which are separated, on average, by
$\sim$550 kpc,  and have diameters of $\sim$1.6 Mpc.  Their masses, as
estimated  by the virial theorem and the projected mass estimator
(Heisler  \etal\ 1985), of $\sim 10^{11.7-13.6}$\msun\ are comparable
to  the mass of the Local Group $\sim 10^{12.4}$\msun\ (Courteau \&
van den Bergh 1999).

\begin{figure}[ht]
\hfill\includegraphics[width=2.3in]{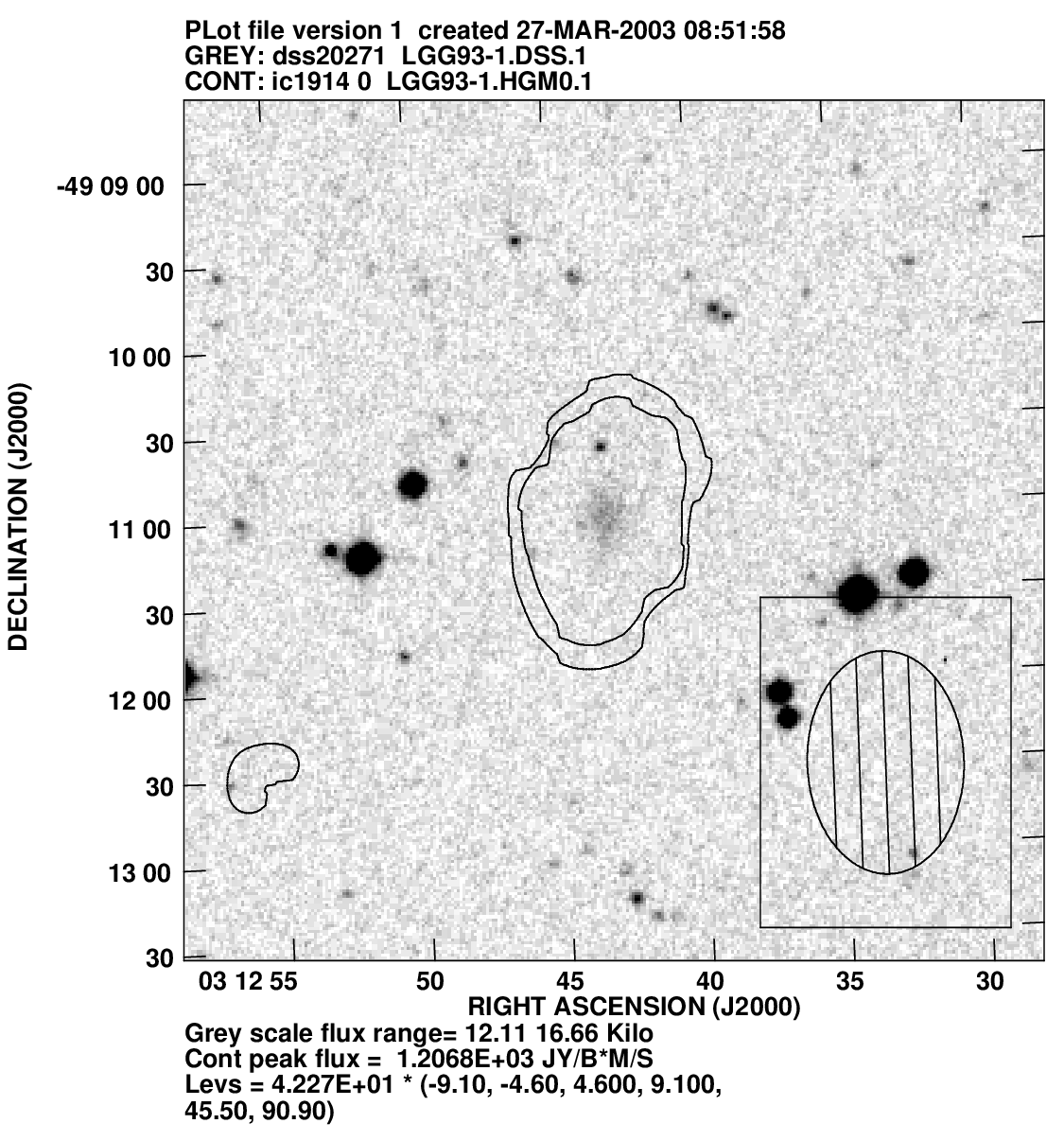}
\hfill\includegraphics[width=2.3in]{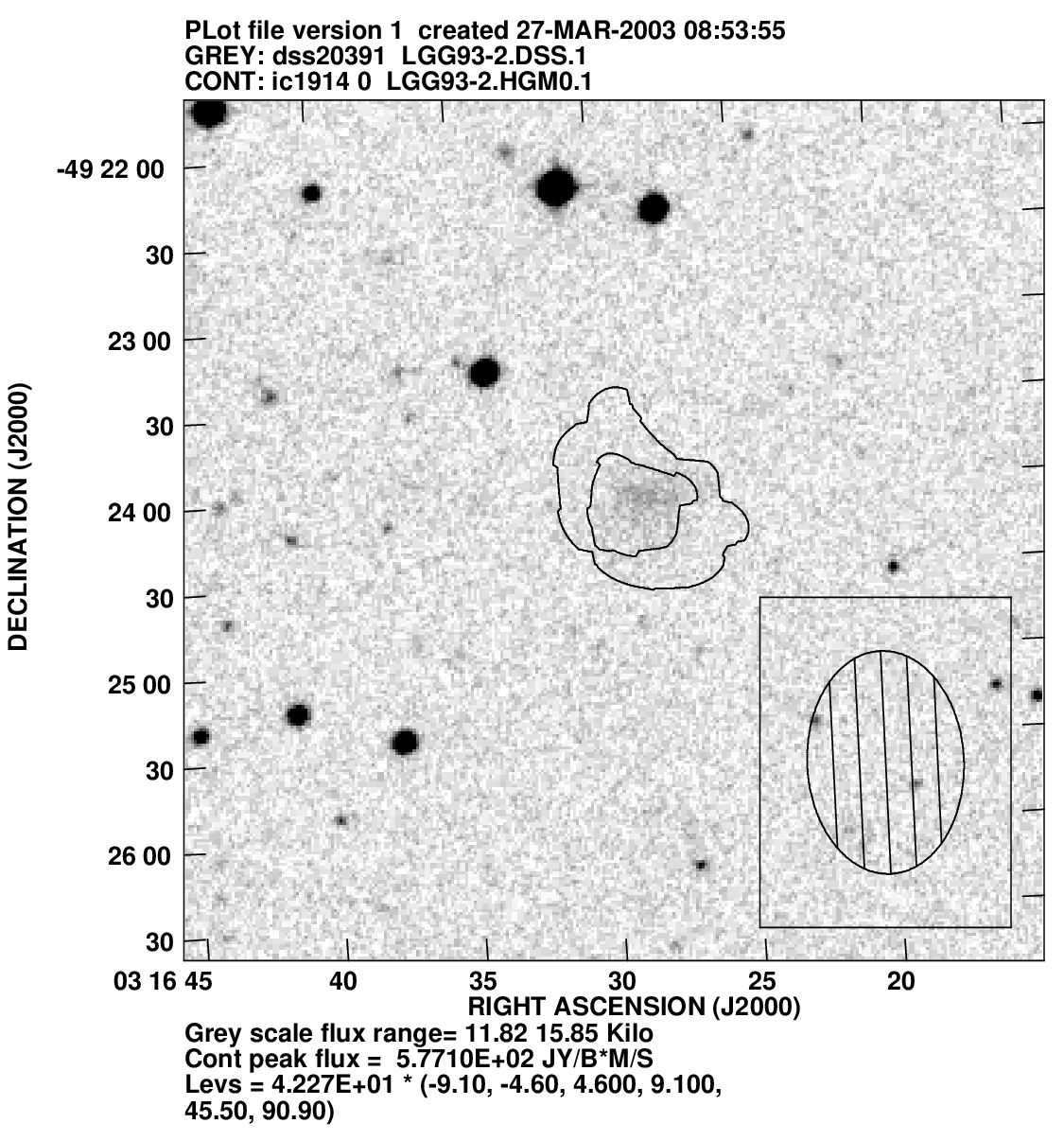}\hspace*{\fill}
\caption{An overlay of the \HI\ total intensity contours on the
optical images of LGG 93-1 (left) and LGG 93-2 (right) also known as
AM 0311-492 and LSBG F200-023, respectively. The ATCA beam is indicated by
the hatched oval in the lower left of both images. \label{fig:opt}}
\end{figure}

We used the Parkes Multibeam and Australia Telescope Compact Array
(ATCA) to survey the entire area of each group down to  a \mhi\
sensitivity of $5-8 \times\ 10^{5}$\msun\ per 3.3 \kms.  All Parkes
detections in the groups were confirmed to be real by the follow-up
ATCA observations.  A total of 64 \HI-rich galaxies were  detected in
the six groups, almost twice the number of optically  cataloged group
galaxies (Garcia 1993) and 50\% more galaxies than  were detected by
HIPASS in the same fields (Meyer et al. 2004).   All of our detections
are associated with optical counterparts and  have properties
consistent with typical spiral, irregular, or dwarf irregular
galaxies.  No analogs to the HVCs were detected.  Examples of two
typical new detections are shown in Figures~\ref{fig:opt}.

\section{Is the Local Group missing dwarf galaxies?}

\begin{figure}[ht]
\hfill\includegraphics[angle=-90,width=2.3in]{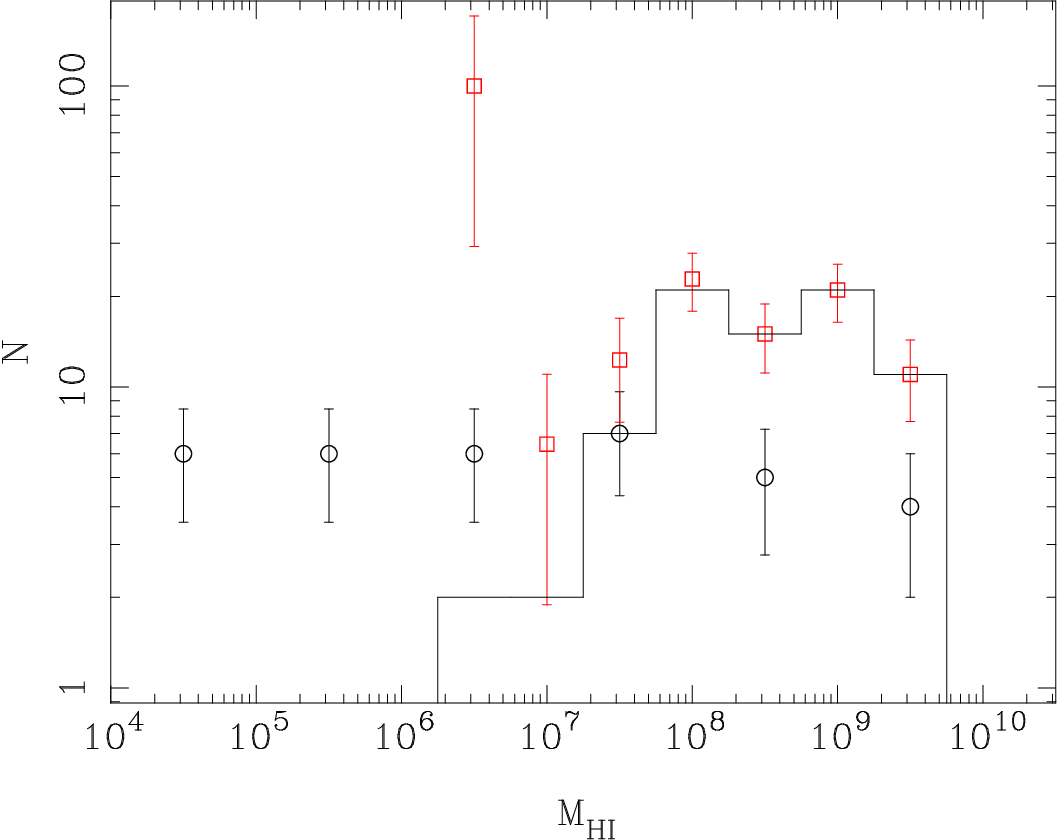}
\hfill\includegraphics[angle=-90,width=2.3in]{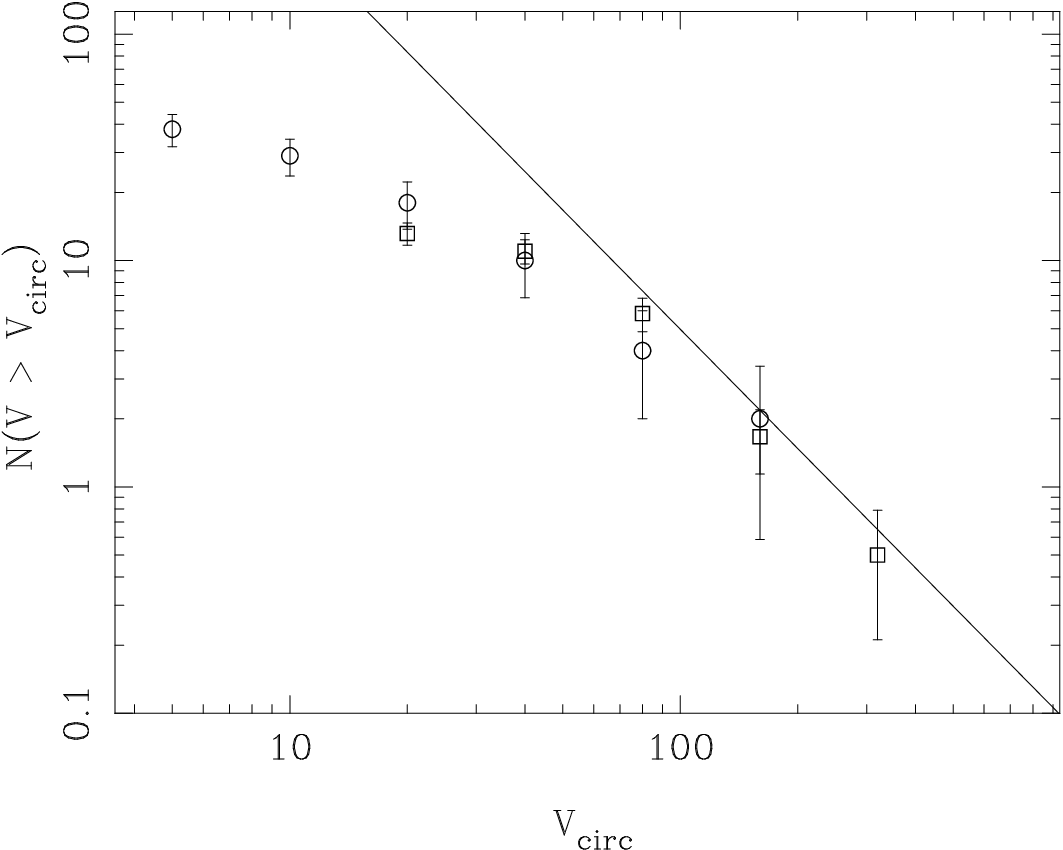}\hspace*{\fill}
\caption{Left:  The \HI\ mass function (\HI MF) of the Local Group
(circles) and the sum of the six loose groups (solid histogram) 
corrected for incompleteness (squares). Right:  The cumulative velocity 
distribution function for the Local Group (circles), and the average 
of the six loose groups (squares).  The solid line represents the CDM model 
of Klypin roughly normalized to the second data point.\label{fig:mf}}
\end{figure}

We wish to know if our sample of loose groups has a significantly
different population of dwarf galaxies as compared to the Local Group
in order to determine if the Local Group is unique in some fashion.
To do this we constructed an \HI\ mass function (\HI\,MF) and a
cumulative circular  velocity distribution function (CVDF), both shown
in Figure~\ref{fig:mf}.   The CVDF is an effective surrogate for a
halo mass function as traced by  luminous matter.  After correcting
for the incompleteness of our survey,  using both the detection rate
of fake sources inserted into our data and by  scaling the HIPASS
completeness function (Zwaan et al. 2004) to our survey  parameters,
we derive a flat \HI\,MF as shown on the left panel of
Figure~\ref{fig:mf} similar to that for the Local Group.  This result
is consistent with the results for the \HI\,MF presented by Zwaan et
al. (2005), who find the \HI\,MF flattens in low density environments
and the results of Tully et al. (2002) who find a similar behavior for
the optical  luminosity function.  Similarly, the six loose groups
have an identical CVDF to the Local Group with both falling below the
predictions of CDM around $V_{circ} \sim 50$\kms.  This demonstrates
that the lack of a large population of dwarf galaxies is not unique to
the Local Group.

\section{Are the High-Velocity Clouds the ``missing satellites''?}

Blitz et al. (1999) and Braun \& Burton (1999) suggested that the HVCs 
seen around the Milky Way may be embedded within the low mass dark 
matter halos that are predicted by CDM models of galaxy formation.
Our non-detection of any HVC analogs in these six loose groups suggests
that if this is the case, then they must be relatively low mass \HI\
clouds, \mhi$\sim 10^{5-6}$\msun, and be within $\sim$150 kpc of the
Milky Way (Pisano et al. 2004).  These constraints are consistent with 
limits found by others through a variety of methods (e.g. Zwaan 2001,
de Heij et al. 2002), but does not necessarily imply that HVCs are 
associated with the CDM dark matter halos (e.g. Maller \& Bullock 2004).  
Deeper \HI\ surveys of individual galaxies are still necessary to better
constrain the nature and origin of HVCs.  



\begin{acknowledgments}
This research was performed while
D.J.P. held a  National Research Council Research Associateship Award
at the Naval Research  Laboratory.  Basic research in astronomy at the
Naval Research  Laboratory is funded by the Office of Naval Research.
D.J.P. also acknowledges generous support from the ATNF via a Bolton
Fellowship and from NSF MPS Distinguished International Research
Fellowship grant AST 0104439.
\end{acknowledgments}



%




\end{document}